\documentclass[prx,twocolumn,superscriptaddress]{revtex4-1}

\usepackage{multirow,eurosym,amssymb,amsfonts,setspace,graphicx,bm,float,amssymb}

\usepackage{color}

\newcommand{\bra}[1]{\langle #1|}
\newcommand{\ket}[1]{|#1\rangle}

\def\be{\begin{equation}}
\def\ee{\end{equation}} 

\def\bsplit{\begin{split}}
\def\nsplit{\end{split}}

\begin{document}
\title{Quantum variational PDE solver with machine learning}

\author{Jaewoo Joo} 
\affiliation{School of Mathematics and Physics, University of Portsmouth, Portsmouth PO1 3QL, UK}

\author{Hyungil Moon} 
\affiliation{Department of Materials, University of Oxford, Parks Road, Oxford OX1 3PH, United Kingdom}

\date{\today}

\begin{abstract}
To solve nonlinear partial differential equations (PDEs) is one of the most common but important tasks in not only basic sciences but also many practical industries. 
We here propose a quantum variational (QuVa) PDE solver with the aid of machine learning (ML) schemes to synergise two emerging technologies in mathematically hard problems. 
The core quantum processing in this solver is to calculate efficiently the expectation value of specially designed quantum operators. 
For a large quantum system, we only obtain data from measurements of few control qubits to avoid the exponential cost in the measurements of the whole quantum system and optimise a pathway to find possible solution sets of the desired PDEs using ML techniques. 
As an example, a few different types of the second-order DEs are examined with randomly chosen samples and a regression method is implemented to chase the best candidates of solution functions with another trial samples. We demonstrated that a three-qubit system successfully follows the pattern of analytical solutions of three different DEs with high fidelity since the variational solutions are given by a necessary condition to obtain the exact solution of the DEs. 
Thus, we believe that final solution candidate sets are efficiently extracted from the QuVa PDE solver with the support of ML techniques and this algorithm could be beneficial to search for the solutions of complex mathematical problems as well as to find good ansatzs for eigenstates in large quantum systems (e.g., for quantum chemistry).
 
\end{abstract}
\maketitle 

\section{Introduction}
\begin{figure}[b]
\centering
\includegraphics[width=1.2\linewidth,trim=5cm 6.5cm 1cm 1.5cm]{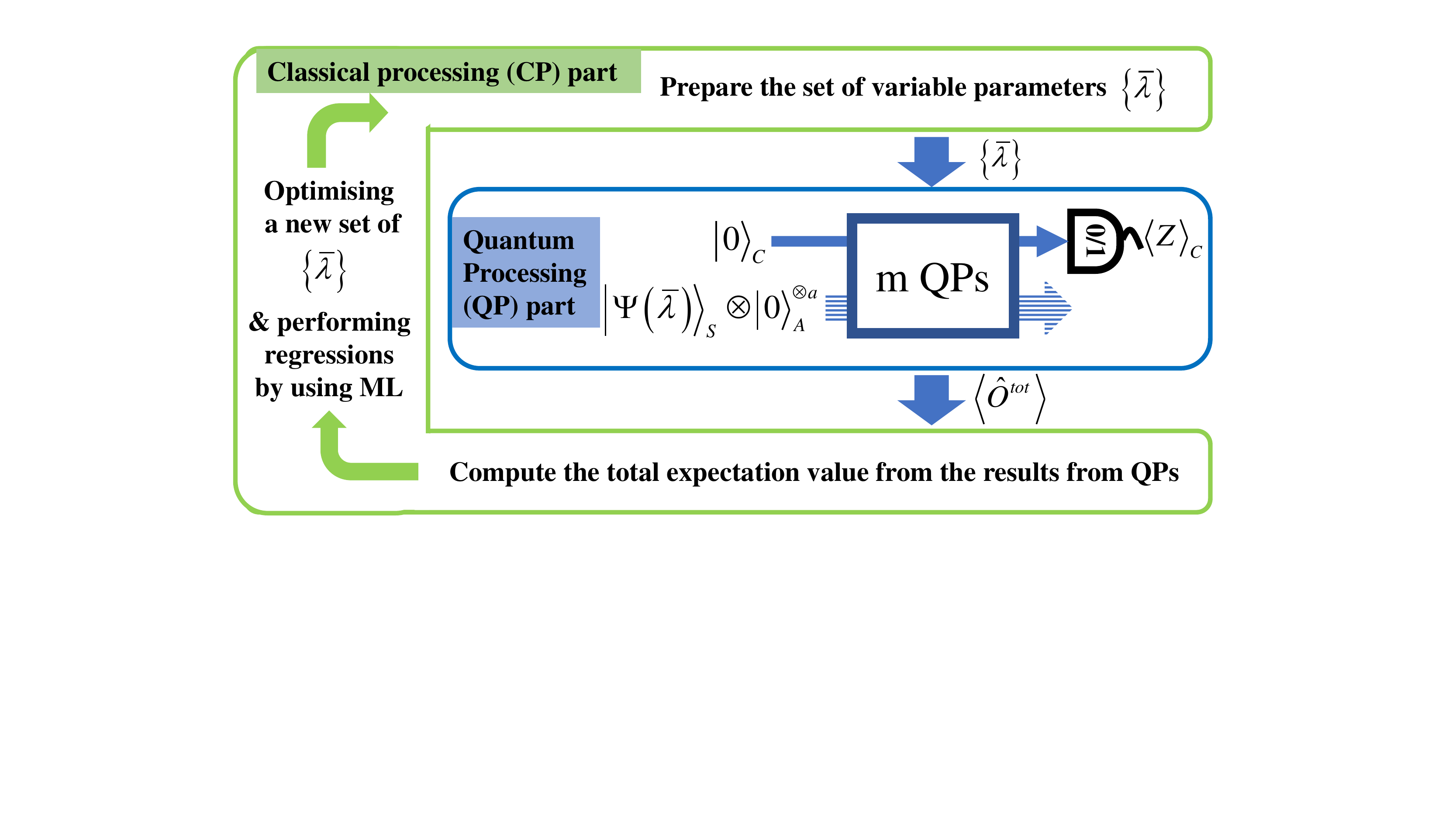}
\caption{Schematics of a quantum variational (QuVa) PDE solver. a set of variational parameters $\{\bar{\lambda}\}$ is optimised by the statistical results from the QP and pre-learned data from the CP by machine learning (ML) schemes.}
\label{fig:intro}
\end{figure}

The investigation of solving a partial differential equation (PDE) has been studied for a long time because these problems ubiquitously occur in not only physics and mathematics but also other sciences, engineering, and technologies. 
To solve a general nonlinear PDE is in common considered as mathematically challenging problems in classical computation when the spanning dimensions are larger. There have been recently a few tactics of how to solve complex PDEs using deep-learning techniques \cite{PDE-DeepLearn}, however, one of the fundamental issues in the classical manner is that it will not be in principle feasible to compute and to memorise the results of very large matrix multiplications in classical processing (CP). For example, in a conventional finite difference method, the adjacent function values should be calculated and recorded in a memory for a number of additions and subtractions of these values to obtain derivative values to solve the PDE while quantum processing (QP) enables to perform a large matrix multiplication at once if quantum coherence is sufficiently maintained in the QP. 

One of the clear advantages in the CP is that it can easily manipulate and perform the optimisation algorithms while it is even a challenging task to design a desired quantum operator, which is corresponding to the target PDE function, only by using the set of one- and two-qubit gates (see Appendix \ref{BasicQI}). To overcome the disadvantage in the QP and to enhance the advantage in the CP, a hybrid (quantum-classical) algorithm might make a synergy between the performance of CP and QP. It is still, however, an open question whether this hybrid computing could fundamentally enhance computational power beyond conventional classical computation or not although several positive efforts have been recently tackled on the problems by quantum-enhanced solvers \cite{AndrewChild, Fontanela, GaussianProc20, Gaitan20, Garcia-Ripoll}. 

We here demonstrate a quantum variational (QuVa) PDE solver coped with machine learning (ML) schemes to overcome a few obstacles faced at the current stage of quantum algorithms. In Fig.~\ref{fig:intro}, the whole algorithm consists of the CP and QP parts and the set of possible solutions are given by iterative CP optimisations with QPs in a variational method. 
For instance, a target PDE is given by ${\cal F}_{tot} \, f (\bar{x})=0$ for equation function ${\cal F}_{tot}$, solution function $ f (\bar{x})$ and dimensional variable set $\{\bar{x}\}$.
The set of variational parameters $\{ \bar{\lambda}\}$ embedded in an ansatz and the information on the target PDE are set at the beginning of the CP part. 

In the QP part of Fig.~\ref{fig:intro} (blue), after the parameters are injected in the QP, the ansatz state $\ket{\Psi}$ in system ($S$), control ($C$) and ancillary qubits ($A$) undergoes the designed QPs. The QP enables to obtain control-qubit measurement outcomes $\langle Z \rangle_C$ and to calculate an expectation value of the total quantum operator $ \hat{\cal{O}}^{tot} $ corresponding to the total PDE function ${\cal F}_{tot}$. The index $m$ denotes the number of QPs given by $\hat{\cal{O}}^{tot} = \sum_{l=1}^{m} \hat{\cal{O}}_l$ and the total expectation value $\langle \hat{\cal{O}}^{tot} \rangle$ is computed by the collection of the QP outcomes. Since the data set of $\langle \hat{\cal{O}}^{tot} \rangle$ is fabricated by the set of the parameters, ML methods are performed to determine the best set of variational parameters $\{ \bar{\lambda}\}$ in order to reduce the computational cost given by many quantum measurements in the QP. Finally, the processes are iteratively performed until we achieve a set of solution candidates to solve the target PDE. 

\section{Part-1 : Theory}
\subsection{Mathematical representation of PDEs}
Let us present a mathematical description of our QuVa PDE solver similar to the well-known method of discretised grid representation. We first define the $j$th-order differential equations (DEs) in a one-dimensional (1D) system with continuous function $ f (x)$ given by $\left( \sum_{i=0}^{j} c_i \left({\partial \over \partial x}\right)^i \right)  f (x) = 0$ for equation coefficient $c_i$ in the $j$-th order. Then, a generalised $j$th-order PDE in $n$ variable dimensions is represented by
\begin{eqnarray}
 && \left( \sum_{k=1}^{n} 
 \sum_{i=0}^{j} c_{i,k} \left({\partial \over \partial x_k}\right)^i \right)  f (x_1,..., x_n) \nonumber \\
 && ~~~~~~~~~~=  {\cal F}_{\partial}  (x_1,..., x_n)\, f (x_1,..., x_n) = 0.~~~~~~~ \label{Eq1}
\end{eqnarray}
If we include a nonlinear effect in this PDE, the nonlinear function operator  ${\cal F}_{NL}$ is inserted in the PDE such as
\begin{eqnarray}
 {\cal F}_{tot} \, f (\bar{x}) = \left( {\cal F}_\partial + {\cal F}_{NL} \right) \, f (\bar{x}) = 0,
 \label{F_tot}
\end{eqnarray}
for the variable dimension set $\{ \bar{x}\}= \{ x_1,..., x_{n}\}$ ($n$ is the dimension number).
In classical simulation, one may use the method of linearlisation to solve the nonlinear PDE. Alternatively, others may numerically find the solution function by an inverse matrix method or by minimising residual value $ Res (\bar{x}) = || {\cal F}_{tot} \, f (\bar{x})||^2  \approx 0$ where $ || \cdot ||$ is called the square norm.

In quantum simulation, the continuous function $f (\bar{x})$ is discretised in a normalised vector form $\ket{\psi (\bar{x})}$ and we here consider only scaled $f(\bar{x})$ in a finite domain due to the limitation of quantum state representations such as
 \begin{eqnarray}
f(\bar{x})  \rightarrow \ket{\psi_N (\bar{x})} = \sum_{g=0}^{2^N-1} {\cal C}_{g} \ket{g},
\end{eqnarray}
where $g$ is represented in either a binary or decimal representation (e.g., $g=10_{(2)} = 2_{(10)}$) and $|{\cal C}_{g}| \le 1$. The quantum system state $ \ket{\psi_N (\bar{x})}$ is made in $N$ qubits, which indicate the number of grid points in the domain. For example, we can use the coefficient ${\cal C}_{g}$ representing the function value $f(x)$ at $x= g/2^N$ in a 1D problem (e.g., in the region of $0\le x \le 1$) and the state $\ket{g}$ is its basis vector at $x= g/2^N$ to form the normalised function vector $\ket{\psi_N (x)}$ with $N$ system qubits \cite{Michael18, Joo19}. 

When the total PDE operator ${\cal F}_{tot}$ in Eq.~(\ref{F_tot}) is reformed in total quantum operator $\hat{{\cal O}}^{tot}$, the total PDE becomes 
\begin{eqnarray}
 \hat{{\cal O}}^{tot} \, \ket{\psi_N (\bar{x})} =
\left( \hat{{\cal O}}_\partial + \hat{{\cal O}}_{NL} \right) \, \ket{\psi_N (\bar{x}) } = \ket{\vec{0}},
\label{General_Quan_PDE01}
\end{eqnarray}
where $\hat{{\cal O}}_\partial$ corresponds to the differential function operator ${\cal F}_\partial$ on $f(\bar{x}) $ and $\hat{{\cal O}}_{NL}$ does to ${\cal F}_{NL}$ ($\ket{\vec{0}}$: null vector).

The core task in the QP is to compute the expectation value $ \bra{\psi_N (\bar{x}) } \hat{{\cal O}}^{tot} \, \ket{\psi_N (\bar{x}) }$ efficiently and the relationship between the expectation value and a residual of PDE is shown as a necessary condition of the PDE solution in Appendix \ref{Append01}. Importantly, even for large $N$, we do not need to measure all the $N$ system qubits directly, which will in general requires exponential cost, but only to preform single control-qubit measurements statistically in each QP to calculate the expectation value of the $2^N \times 2^N$ matrix $\hat{\cal O}^{tot}$ \cite{Ekert_Oi} (see details in Appendix \ref{General-expectation-calculation}). Note that the number of variational parameters are commonly different from the equation dimension $n$ because a 1D system can be investigated by several variational parameters. The following subsections describe the method of mapping the mathematical derivatives to quantum operators.

\subsection{Expectation value of quantum subtractor $\hat{A}^{\dag}$ for derivative operations}
We now describe how to calculate the expectation value of a specific quantum operator named a quantum subtractor $\hat{A}^{\dag}$ equivalent to operator $\hat{A}$ also known as a quantum adder operator \cite{Adder}. The quantum subtractor (adder) mimics a shifting operation on the system qubits $ \ket{\psi(x)} = \hat{A}^{\dag}  \ket{\psi(x+\delta L)}$ ($ \ket{\psi(x)} = \hat{A} \ket{\psi(x-\delta L)}$) for all the values of the function vector $ \ket{\psi(x)}$ where $\delta L$ is the unit grid space. 
Then, it has been known that the statistics of single-qubit measurements in a control qubit brings the expectation value of the quantum operator embeded on system qubits \cite{Ekert_Oi,OxfordError-mittigation,OxfordError-mittigation2} (see the details in Appendix \ref{General-expectation-calculation}).

It is mathematically true that the translation operator with $\delta L$ is given by $\hat{A}^{\dag}$ on $\ket{\psi(x)}$  is defined by
\begin{eqnarray}
\hat{A}^{\dag} \ket{\psi(x)} = e^{-(\delta L)\, \partial_x}\, \ket{\psi(x)}  = \ket{\psi(x-\delta L)} \,. \nonumber \\
\label{eq:Translate01}
\end{eqnarray}
A periodic boundary condition is applied here in normalised $\ket{\psi({x})}$ implies $\ket{\psi(0)} = \ket{\psi(1)}$ for $0\le x \le 1$ because we utilise an unitary operator to describe the quantum subtractor in the domain. In fact, the ideal translation operator could be implemented beyond the finite regions in a specific physical system as a non-unitary gate (e.g., a photon shift operation \cite{My-amp-paper}).

\subsubsection{The 1st and 2nd derivative quantum operators}
To calculate the expectation value of the quantum derivative operators, we adopt the concept of the finite difference method to represent quantum operators. More precisely, one can define the second derivative quantum operator given by
\begin{eqnarray}
\hat{\cal O}_{\partial^2} = {1 \over (\delta L)^2} \left( \hat{A} + \hat{A}^{\dag} -2 \hat{I} \right),
\label{eq:2ndDerivative01}
\end{eqnarray}
since the second-order (2O) differential form is approximately given by
\begin{eqnarray}
 {\partial^2 \over \partial x^2} \ket{\psi(x)} && \, \approx  {1 \over (\delta L)^2} \left( \hat{A} + \hat{A}^{\dag} -2 \hat{I} \right) \ket{\psi(x)} = \hat{\cal O}_{\partial^2} \ket{\psi(x)}, \nonumber \\
\label{eq:2ndDerivative02}
\end{eqnarray}
where $\hat{I}$ is an identity operator. Similarly,
the first derivative form is defined by
\begin{eqnarray}
{\partial \over \partial x} \ket{\psi(x)} &&\, \approx {1 \over  \delta L} \Big( { \hat{I} -  \hat{A}^{\dag}} \Big) \ket{\psi(x)} = \hat{\cal O}_{\partial^1}  \ket{\psi(x)}, 
\label{eq:1stDerivative02}
\end{eqnarray} 
with the first-order derivative quantum operator
\begin{eqnarray}
\hat{\cal O}_{\partial^1} ={1 \over  \delta L} \Big( { \hat{I} -  \hat{A}^{\dag}} \Big) \, .
\label{eq:1stDerivative01}
\end{eqnarray}

For example, if a typical 2O DE with function $f (x)$ is represented by
\begin{eqnarray}
\left( \kappa_2 {\partial^2 \over \partial x^2} + \kappa_1 {\partial \over \partial x} + \kappa_0 \right) f (x) =0,
\label{eq:2-order-PDE01}
\end{eqnarray}
for constants $\kappa_j$ ($j=0,1,2$), its discretised version is given by
\begin{eqnarray}
\hat{\cal O}_{\partial}  \ket{\psi(x)} &=& \left( \kappa_2 \hat{\cal O}_{\partial^2}+ \kappa_1 \hat{\cal O}_{\partial^1} + \kappa_0 \hat{I} \right) \ket{\psi(x)}  = \ket{\vec{0}}.~~~~~~~
\label{eq:2-order-PDE02}
\end{eqnarray}

\subsubsection{Expectation values of the derivative quantum operators}
For a 1D system, we calculate the expectation value of the first derivative operator $\hat{\cal O}_{\partial^1}$ and the second one $\hat{\cal O}_{\partial^2}$ 
\begin{eqnarray}
\langle \hat{\cal O}_{\partial^1} \rangle &&  \approx  { 1 \over \delta L} \Big( 1 -  \Re \left[\left< \hat{A}^{\dag} \right> \right]  - \Im \left[\left< \hat{A}^{\dag} \right> \right]  \Big),
\label{eq:Expect01} \\
\langle \hat{\cal O}_{\partial^2} \rangle && \approx  {2 \over \, (\delta L)^2} \left( \Re\left[\left< \hat{A}^{\dag} \right> \right]  - 1 \right),
\label{eq:Expect02}
\end{eqnarray}
where $\Im \left[ \, \right] $ and $\Re \left[ \,  \right] $ are the imaginary and real part of the expectation value.

Thus, a general form of 2O DEs is given with $\hat{\cal O}_{\partial} =  \kappa_2 \hat{\cal O}_{\partial^2} + \kappa_1 \hat{\cal O}_{\partial^1} + \kappa_0 \hat{I}$ and its expectation value is given by
\begin{eqnarray}
\langle \hat{\cal O}_{\partial} \rangle && = \left( {2 \, \kappa_2 \over \, (\delta L)^2} - { \kappa_1 \over \delta L} \right) \, \Re\left[\left< \hat{A}^{\dag} \right> \right]   \nonumber \\ &&~~~ - { \kappa_1 \over \delta L} \, \Im \left[\left< \hat{A}^{\dag} \right> \right]  + \kappa_0 + { \kappa_1 \over \delta L} - {2 \, \kappa_2 \over \, (\delta L)^2} .
\label{eq:Expect03}
\end{eqnarray}
In general, the higher-order derivatives are feasible in a similar approach using multiplying quantum adder and subtractor operators and it could be fit to the research area of relativistic quantum mechanics with high-order momentum operators.

\subsection{Higher-dimensional PDE}
For a multi-dimensional system, we expand the concept of the expectation values with multi-dimenson set $\{\bar{x}\}$ such as $\langle \hat{\cal O}_{\partial} (\bar{x}) \rangle$. 
As an example in a 2D system, the first- and second-order derivative operators are given by
\begin{eqnarray} 
\hat{\cal O}_{\partial^2} (\bar{x}) &&  = \hat{\cal O}_{\partial^2} (x) \otimes \hat{I} (y)  + \hat{I}(x) \otimes  \hat{\cal O}_{\partial^2} (y) ,
\label{eq:2DLaplace01} \\
\hat{\cal O}_{\partial^1} (\bar{x}) && = \hat{\cal O}_{\partial^1} (x) \otimes \hat{I} (y)  + \hat{I}(x) \otimes  \hat{\cal O}_{\partial^1} (y) ,
\label{eq:2DLaplace02}
\end{eqnarray}
where $\delta x = \delta y = \delta L$ for ${\partial^2 / \partial x^2} + {\partial^2 / \partial y^2} \approx \hat{\cal O}_{\partial^2} (x,y)$ and  ${\partial / \partial x} + {\partial / \partial y} \approx \hat{\cal O}_{\partial^1} (x,y)$ for $\{\bar{x}\} = \{x,y\}$.
Thus, if we focus on the separation of variables in separable ansatz state $\ket{\Psi(\bar{x})} = \ket{\psi(x)} \otimes  \ket{\phi(y)}$, the total expectation value for $\ket{\Psi(\bar{x})} $ is given by 
\begin{eqnarray}
\langle \hat{\cal O}_{\partial} (\bar{x}) \rangle &&= \langle \hat{\cal O}_{\partial} (x) \rangle  + \langle \hat{\cal O}_{\partial} (y) \rangle  \nonumber \\
&& = \sum_{j=x,y} \langle \kappa_{j2} \, \hat{\cal O}_{\partial^2} (j)\, + \kappa_{j1} \, \hat{\cal O}_{\partial^1} (j) \, + \kappa_{j0} \, \hat{I} \rangle,~~~~~~
\label{eq:2DExpect02}
\end{eqnarray}
in order to solve 2O PDEs. Thus, we enable to find the expectation values for each term independently and to add all together after the QP part.
We will give more concrete examples in different 1D DEs and discuss the details of how to deal with the nonilnear and other operators in the following section.

\section{Part-2: Differential equations in a 1D system}
It is very crucial to choose an ansatz state in variational methods because it describes a trial function with a set of variational parameters $\{\bar{\lambda}\}$ in $N$ qubits $\ket{\psi^d_N (\bar{\lambda})}$ also called an ansatz system state with depth $d$ (see Appendix \ref{N-qubit ansatz states}). Ref.~\cite{Garcia-Ripoll} proposed a few types of ansatz states and we chose a real-value ansatz state with $\hat{R}^Y$ gates for ansatz states $\ket{\psi^{d}_N (x)}$, which implies that all the elements can cover positive and negative real values for real-value function $f(x)$. Note that $d$ represents the depth of quantum circuits to create the system state with $N$ qubits. This ansatz approach is commonly utilised as one of the excellent candidates to represent states for quantum chemistry in superconducting circuits \cite{IBM_QuanChem} and trapped ions \cite{Trap-ion}.

\subsection{Simple 1D DEs with ${\kappa}_2 = 1$}
Let us first fix ${\kappa}_2 = 1$ and ${\kappa}_1 = 0$ for the simplest DEs in Eq.~(\ref{eq:2-order-PDE01}) and this equation is called the Helmholtz equation given by $ \left( \partial^2 / \partial x^2 +  \kappa_0 \right) \, f (x) = 0$. We consider that $ f (x)$ and $\ket{\psi^d_N (x)}$ are periodic real functions with an end-point restriction (e.g., $f (0) = f (1)$ and $\ket{\psi^d_N (0)} = \ket{\psi^d_N (1)}$). For the ansatz state $\ket{\psi^d_N (x)}$, the expectation value of the derivative operator $ \hat{\cal O}_{\partial} $ is simplified by
\begin{eqnarray}
\langle \hat{\cal O}_{\partial} \rangle &&= {2 \, \over \, (\delta L)^2} \, \Re \big[ \langle \hat{A}^{\dag} \rangle \big] + \kappa_0 - {2 \over \, (\delta L)^2} ,~~~
\label{eq:Expect04}
\end{eqnarray}
and we only need to calculate $\Re \big[ \langle \hat{A}^{\dag} \rangle \big]$ produced by measuring the control qubit statistically in the QP.
For given $\kappa_0$ and $\delta L$, the statistical data of the control qubit measurement shows the landscape of $\Re \left[ \langle  \hat{A}^{\dag} \rangle \right] $ with respect to the parameter set $\{\bar{\lambda}\}$.

For a 2O DE given in Eq.~(\ref{eq:Expect03}), the total expectation value with $\kappa_2=1$ is similarly given by
\begin{eqnarray}
\langle \hat{\cal O}_{\partial}  \rangle && = {1 \over \delta L} \left( {2  \over \delta L} - { \kappa_1} \right) \,\Re \big[ \langle \hat{A}^{\dag} \rangle \big] + \kappa_0 + { \kappa_1 \over \delta L} - {2 \over \, (\delta L)^2} .~~~~~~
\label{eq:Expect05}
\end{eqnarray}
Note that $ \Im \big[ \langle \hat{A}^{\dag} \rangle \big]$ vanished due to real-value ansatz $\ket{\psi^d_N (x)}$ and real number elements in $\hat{A}^{\dag}$. 
From Eq.~(\ref{eq:Expect05}), it shows that $\kappa_1$ and $\delta L$ change the scale of $ \langle \hat{A}^{\dag} \rangle $ and the rest part makes a shift of the landscape $\langle \hat{\cal O}_{\partial}  \rangle$. 
Therefore, the case of this linear 2O DE requires the expectation value of $\hat{A}^{\dag}$ with fixed $\kappa_0$, $\kappa_1$ and $\delta L$ in the algorithm.

\subsection{Generalised second-order nonlinear DEs}
There are a variety of the nonlinear function operator ${\cal F}_{NL}$ in Eq.~(\ref{F_tot}), which could be represented by its quantum operator $\hat{\cal O}_{NL}$ once one finds how to implement it in quantum circuits. One of the well-known nonlinear operators is a self-interaction nonlinear term  ${\cal F}_{NL} = \kappa_n | f ({x})|^2$ with nonlinearity strength $\kappa_n$ in Gross-Pitaevskii (GP) equation and is inspired by the quantum phenomena in Bose-Einstein condensation \cite{GP-equation}. Then, a generalised 2O nonlinear DE with ${\kappa}_2 = 1$ is written with general potential $V(x)$ by
\begin{eqnarray}
\left( {\partial^2 \over \partial x^2} \, + {\kappa}_1 {\partial \over \partial x} + \kappa_0 + V(x) + \kappa_{n}  \left| f (x)\right|^2 \right) f (x) = 0 \, .~~~~~
\label{eq:2NDE02}
\end{eqnarray}

Let us first consider the expectation value of this nonlinear DE with $\ket{\psi^d_N (x)}$ given by
\begin{eqnarray}
\langle \hat{\cal O}^{tot} \, \rangle = \bra{\psi^d_N (x)} \left(  \hat{\cal O}_{\partial} + \hat{\cal O}_{NL} \right) \ket{\psi^d_N (x)} ,
\label{eq:2-order-PDE06}
\end{eqnarray}
where $ \hat{\cal O}_{NL} = \hat{\cal V} + {\kappa}_{n}  \, \hat{\rho}_D $ for potential part $\hat{\cal V} = \sum_{g=0}^{2^N-1} V_{g} \, \ket{g}\bra{g}$ and GP-type interaction $\hat{\rho}_D = \sum_{g =0}^{2^N-1} \left| {\cal C}^d_g \right|^2 \ket{g} \bra{g} $.
Note that $V_g$ indicates the height of the potential at $x=g/2^N$. Thus, $\hat{\cal O}_{NL}$ is explicitly represented by the diagonal matrix
\begin{eqnarray}
\hat{\cal O}_{NL} = \sum_{g=0}^{2^N-1}  \left( V_{g} +{\kappa}_{n} \left| {\cal C}^d_g  \right|^2 \right) \ket{g}\bra{g},
\label{eq:F_N}
\end{eqnarray}
and then the expectation value of $\hat{\cal O}_{NL}$ with $\ket{\psi^d_N (x)}$ is written by
\begin{eqnarray}
\langle \hat{\cal O}_{NL} \rangle &&=  \langle \hat{\cal V} \rangle + {\kappa}_{n} \langle \hat{\rho}_D \rangle \nonumber \\
&&= \sum_{g =0}^{2^N-1} V_{g} \left| {\cal C}^d_g \right|^2 + {\kappa}_{n} \sum_{g =0}^{2^N-1} \left| {\cal C}^d_g  \right|^4.
\label{eq:Expect-F_N}
\end{eqnarray}
Note that it is generally difficult to compute the second (nonlinear) term if we do not deal with all the elements of $\ket{\psi^d_N (x)}$ but do with the set of variational parameters.

\begin{figure}[b]
\centering
\includegraphics[width=0.95\linewidth,trim=2cm 8cm 2cm 0cm]{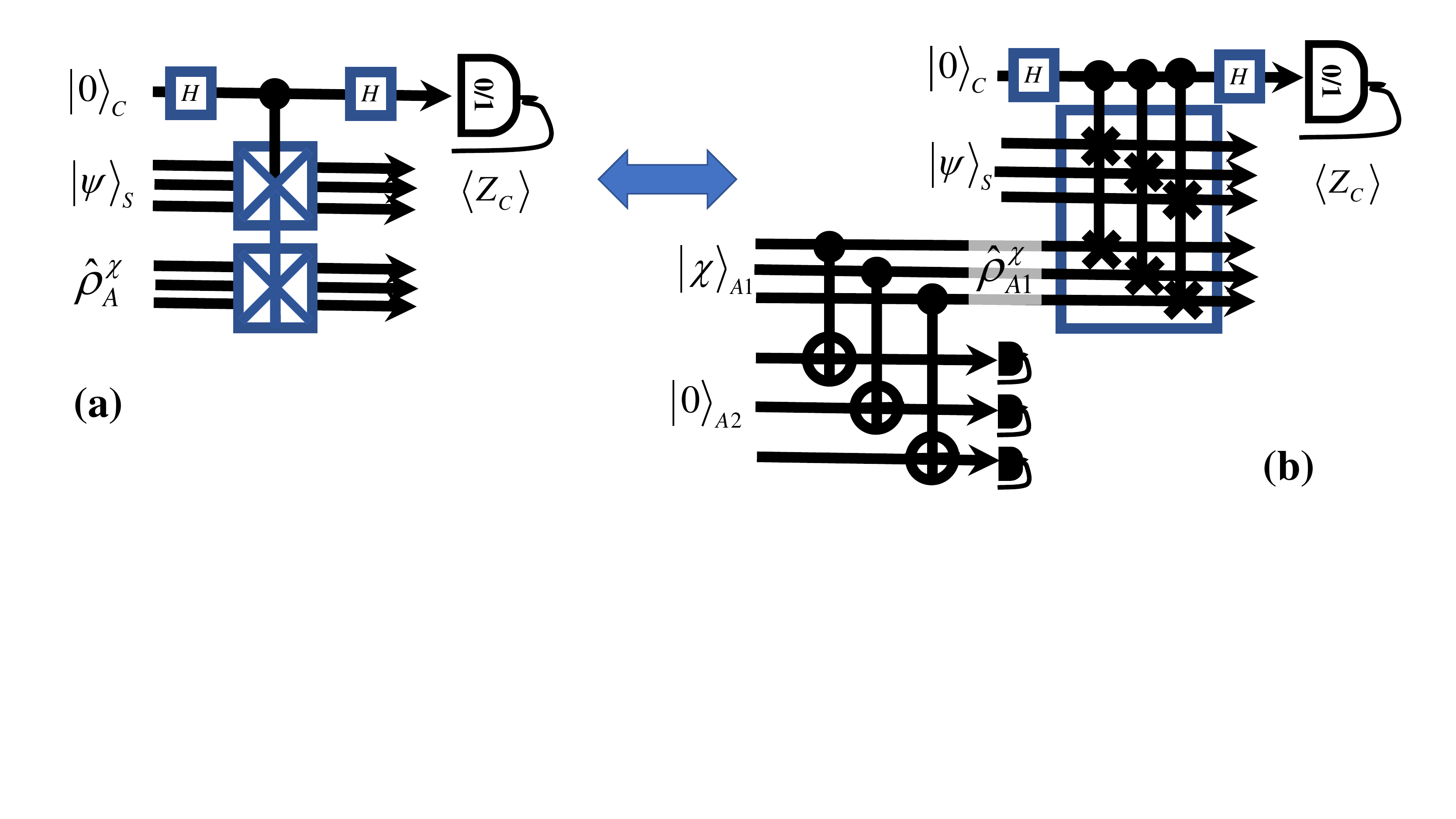}
\caption{ Two equivalent quantum circuits for calculating the expectation value of $\hat{\rho}^{\chi}$ using a block-SWAP gate. (a) the desired mixed state $\hat{\rho}^{\chi}= \sum_g \chi_g \ket{g}\bra{g}$ is injected in $B$ while the mixed state $\hat{\rho}^{\chi}$ can be also made from pure state $\ket{\chi} = \sum_g \sqrt{\chi_g} \ket{g}$ prepared in $S_2$ if ancillary qubits are used through the artificial decoherence mechanism in (b).
}
\label{fig:03}
\end{figure}

It is important to notify that the forms of $\hat{\cal V}$ and $\hat{\rho}_D$ are a $2^N \times 2^N$ diagonal matrices and can be represented by mixed states (e.g., $\hat{\rho}^{\chi}$ in Fig.~\ref{fig:03}(a)).
Thus, we enable to build the expectation value calculator to obtain $\langle \hat{\cal V} \rangle$ and $\langle \hat{\cal \rho}_D \rangle$ through a controlled-SWAP quantum circuit (see details in Appendix \ref{General-expectation-calculation} and \ref{Overlap_fidelity}). 

Based on Eqs.~(\ref{eq:Expect05}) and (\ref{eq:Expect-F_N}), the total expectation value is given by
\begin{eqnarray}
\langle \hat{\cal O}^{tot} \, \rangle &=& {1 \over \delta L} \left( {2  \over \delta L} - { \kappa_1} \right) \,\Re \big[ \langle \hat{A}^{\dag} \rangle \big]  +  \langle \hat{\cal V} \rangle \nonumber \\
&& + {\kappa}_{n} \langle \hat{\rho}_D \rangle + \kappa_0 + { \kappa_1 \over \delta L} - {2 \over \, (\delta L)^2} .
\label{eq:Genreal_total_O}
\end{eqnarray}
We would like to emphasise that it is also feasible to investigate beyond the GP-type nonlinear term if we create the copies of the system state and extra controlled unitary gates in the QP \cite{Joo19}.

\subsection{How to optimise the set of parameters with ML}
Gaussian process regression (GPR) is a non-parametric Bayesian regression method. It is well known for providing reliable uncertainty estimates of a regression target. The uncertainty estimates can be used to design sample efficient optimisation algorithms, called Bayesian optimisation and root-finding algorithms as well~\cite{Bect12,Bichon}. A recent study shows that Gaussian process can be efficiently used to find a many-body entangled state as a ground state \cite{GaussianProc20}. In our method, GPR is used for sequentially collecting data, whereas Ref.~\cite{GaussianProc20} focuses on a new representation of the Gaussian process and its states.

The goal of our QuVa PDE solver is to find candidates $\lbrace \bar{\lambda} \rbrace$ that makes $\langle \hat{Q}^{tot} \rangle$ close to 0, which is essentially a multi-dimensional root-finding problem. 
First, we set up the initial variational parameter set $\{ \bar{\lambda}\}$ randomly with fixed input parameters ($d$,  $\kappa_2$, $\kappa_1$, $\kappa_0$,  $V_g$ and $\kappa_n$) in the CP. Second, we iteratively obtain the the expectation values ($\langle \hat{A}^{\dag} \rangle$, $\langle \hat{\rho}^{\chi} \rangle$ and $\langle \hat{\rho}_D \rangle$) from the QPs and calculate $ \langle \hat{\cal O}^{tot} \rangle$ for each $ \ket{\psi^d_N (x)}$ given by the variational set in depth $d$. 
From the individual outcomes of $ \langle \hat{\cal O}^{tot}\rangle$, we then estimate the best next parameters $\{ \bar{\lambda}\}$, which maximises the acquisition function for the root-finding problem. The acquisition function evaluates the value of measuring $\langle \hat{O}^{tot} \rangle$ at given $\{ \bar{\lambda}\}$. There are two factors that makes the acquisition value high at the given $\{ \bar{\lambda}\}$. The first factor is when the GPR model estimates $\langle \hat{O}^{tot} \rangle$ close to zero, and the other factor is when the uncertainty estimate from the GPR model is high. The two factors has an exploration-exploitation trade-off, and the acquisition function balances the two factors \cite{Bect12,Bichon}.

\section{Part-3: Demonstration for 1D nonlinear DE}

\subsection{Expectation values with $ \ket{\psi^d_3 (x)}$}
We demonstrate three examples of how to obtain the solution candidate sets of DEs using a three-qubit system state.
For example, eight point values of function $f(x)$ for $0\le x \le 1$ can be represented in the system qubits 
\begin{eqnarray}
f(x) \approx \ket{\psi_3 (x)} && = \sum_{g=0}^{7} {\cal C}_{g}\, \ket{g}.
\end{eqnarray}
It implies that the coefficient ${\cal C}_{000}$ represents the function value $f(x=0)$, ${\cal C}_{001}$ does $f(x=1/8)$, ${\cal C}_{010}$ does $f(x=1/4)$ and so on until ${\cal C}_{111}$ does $f(x=7/8)$ while we keep a periodic boundary condition such as $f(0) = f(1)$. As shown in Appendix \ref{N-qubit ansatz states}, the variational ansatz $\ket{\psi^d_3 (x)}$ with depth $d$ contains 8 function values with a specific parameter set $\{d, \bar{\lambda} \}$ and is written by
\begin{eqnarray}
\ket{\psi^d_3 (x)} && = \ket{\psi^d_3 (\bar{\lambda})} = \sum_{g =0}^{7} {\cal C}^d_g  (\bar{\lambda}) \ket{g}  = \sum_{jkl =0}^{1} {\cal C}^d_{jkl}  (\bar{\lambda})\, \ket{j\, k\, l}, \nonumber \\
&& 
\label{eq:Sol_01}
\end{eqnarray}
where fixed $g$ represents fixed location $x$ and ${\cal C}^d_{g}$ with fixed $\bar{\lambda}$ does the function value $f(x)$ at $x=g/8$.

Based on Eq.~(\ref{eq:Expect04}) with $\delta L=1/(2^3) = 1/8$, the expectation value of the Helmholtz operator becomes
\begin{eqnarray}
\langle \hat{\cal O}^{tot} (\bar{\lambda}) \rangle  = 128 \, \Re \left[ \langle  \hat{A}^{\dag} \rangle \right]  + (\kappa_0 - 128), 
\label{eq:Helm01}
\end{eqnarray}
and the solution condidate set is achieved by the ansatz states under the condition of $\langle \hat{\cal O}^{tot} (\bar{\lambda}) \rangle \approx 0$ from the data of $\Re \left[ \langle  \hat{A}^{\dag} \rangle \right] $. 
Similarly, the expectation value of the 2O DE in Eq.~(\ref{eq:Expect05}) is equal to 
\begin{eqnarray}
\langle \hat{\cal O}^{tot} (\bar{\lambda}) \rangle  = {8} \left( {16} - { \kappa_1} \right) \, \Re \left[ \langle  \hat{A}^{\dag} \rangle \right] + \left( \kappa_0 + { 8 \, \kappa_1  } - 128 \right). \nonumber \\
\label{eq:Helm02}
\end{eqnarray}
For the condition of  $\kappa_2 = 1$ and $\kappa_1 = 16$, it is very unlikely to be solved using the three-qubit ansatz states because $\langle \hat{\cal O}^{tot} \rangle$ is then independent from $ \langle \hat{A}^{\dag} \rangle $.
Thus, for this DE with $\kappa_1 \approx 16$, we need more system qubits due to the condition of $\kappa_1 \approx 2 \kappa_2 / \delta L$ in Eq.~(\ref{eq:Expect03}).

For the generalised 2O DE, we assume a potential operator $\hat{\cal V}$ as a harmonic potential, which is proportional to $V_{max} (1-2x)^2$ with maximum height $V_{max}$ at $x=0,1$. Then, the potential operator for three qubits is given by
\begin{eqnarray}
\hat{\cal V} && = \sum_{g=0}^7 V_g \ket{g} \bra{g} = {4\,V_{max}\over 11} \sum_{g=0}^7 \left(1- {g\over 4} \right)^2  \ket{g} \bra{g},
\label{eq:Potential01}
\end{eqnarray}
where $x=g/8$ and the maximum potential height $V_{max}$. 
In Fig.~\ref{fig:03}(a), the desired mixed state $\hat{\rho}^{\chi}$ might be directly built in $A$ such as $\hat{\rho}^{\chi} = {\hat{\cal V} /  V_{max}} $. 
If we put $\ket{\psi^d_3 (x)}$ in system mode $S$ and $\hat{\rho}^{\chi}$ in $A$, the expectation value of the potential operator is represented by
\begin{eqnarray}
\langle \hat{\cal V} \rangle && = V_{max} \langle \hat{\rho}^{\chi} \rangle. 
\label{eq:Potential03}
\end{eqnarray}
As shown in Fig.~\ref{fig:03}(b), the operator $\hat{\cal{V}}$ is alternatively implemented by the specific three-qubit state given by
\begin{eqnarray}
\ket{\chi_3}_{A1} =  {1 \over 2\sqrt{11}}  && \Big( 4 \ket{000} + 3 \ket{001}  + 2 \ket{010} \nonumber \\
&& +  \ket{011} + \ket{101} + 2 \ket{110}+ 3 \ket{111} \Big),~~~~
\label{eq:Potential02}
\end{eqnarray}
by using a quantum circuit through the artificial decoherence (see details in Appendix \ref{Arti-decoh}). 

For the nonlinear part, if we prepare $\ket{\psi^d_3 (x)}$ in both modes ($S$ and $A1$) in Fig.~\ref{fig:03}(b), the state in mode $A1$ turns into the mixed state given by $\hat{\rho}_D =\sum_{g=0}^7 |{\cal C}^d_g|^2 \ket{g}\bra{g}$, and the expectation value is given by 
\begin{eqnarray}
\langle \hat{\rho}_D \rangle && = \sum_{g=0}^7 \left| {\cal C}^d_g \right|^4. 
\label{eq:NL01}
\end{eqnarray}
Therefore, the three expectation values ($\langle \hat{A}^{\dag} \rangle$, $\langle \hat{\rho}^{\chi} \rangle$ and $\langle \hat{\rho}_D \rangle$) are individually calculated through the QPs and the total expectation value with a three-qubit system is given by
\begin{eqnarray}
\langle \hat{\cal O}^{tot}  (\bar{\lambda}) \rangle  &=&  8 \left(  16 -   \kappa_1 \right) \,  \Re \left[ \langle \hat{A}^{\dag} \rangle \right] \nonumber \\
&& +  \left( \kappa_0 + 8 \kappa_1 - 128 \right)  + V_{max} \langle \hat{\rho}^{\chi} \rangle + {\kappa}_{n} \langle \hat{\rho}_D \rangle, \nonumber \\
&&
\label{eq:Demon01}
\end{eqnarray}
for the nonlinear strength ${\kappa}_{n}$.

{
\begin{widetext}

\begin{figure}[t]
\centering
\includegraphics[width=0.45\linewidth,trim=1cm 0cm 0cm 0cm]{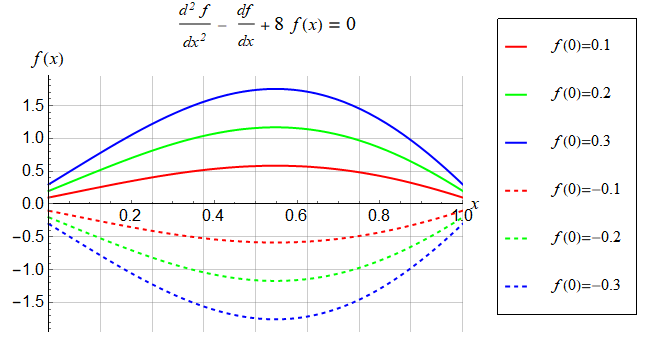}
\includegraphics[width=0.48\linewidth,trim=0cm 0cm 1cm 0cm]{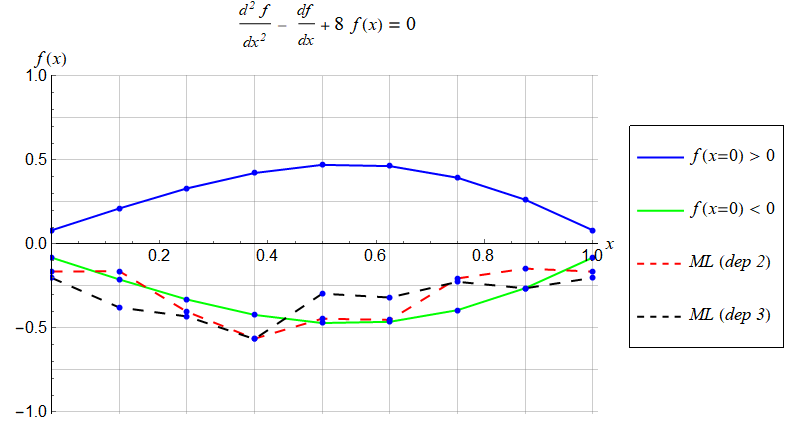}
\includegraphics[width=0.45\linewidth,trim=1cm 0cm 0cm 0cm]{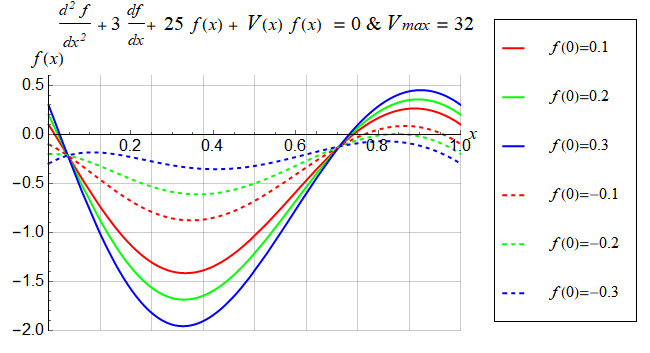}
\includegraphics[width=0.48\linewidth,trim=0cm 0cm 1cm 0cm]{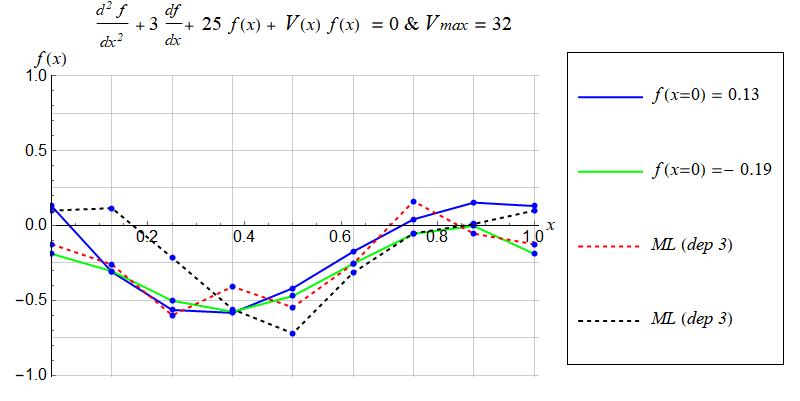}
\includegraphics[width=0.45\linewidth,trim=1cm 0cm 0cm 0cm]{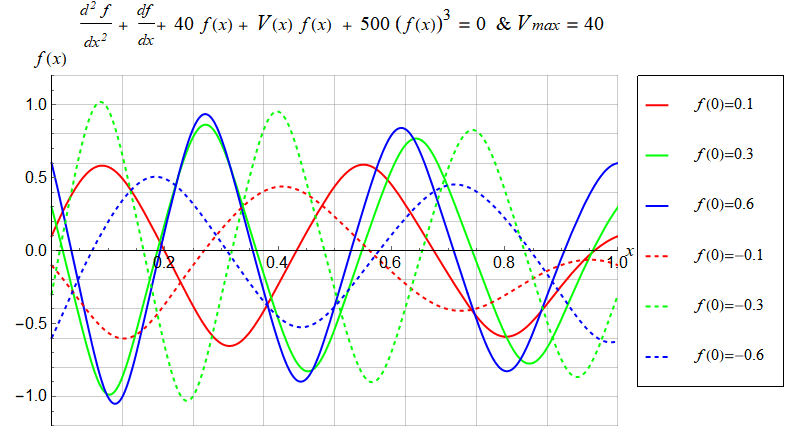}
\includegraphics[width=0.48\linewidth,trim=0cm 0cm 1cm 0cm]{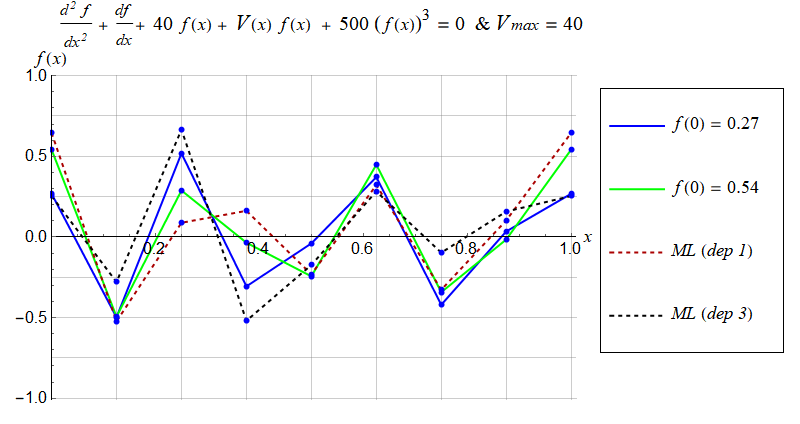}
\caption{Analytic and ML solutions for 1D 2O DE. The DE parameters are $\{ \kappa_2, \kappa_1,\kappa_0, V_{max}\} =  \{1, -1, 8, 0 \}$ (top),  $\{1, 3,25, 32\}$ (middle) and  $\{1, 1, 40, 40\}$ with $\kappa_n = 500$ (bottom). Analytic solutions are depicted depending on the initial function value $f(0)$ in $0\le x \le 1$ with a periodic boundary condition $f(0)=f(1)$. To compare with discretised quantum solutions, the solid lines are given as normalised functions with 8 points based on the analytical solutions.On the right side, the best ML-aid solutions in red dashed lines approximately gives fidelity 0.92 with $\ket{\psi^2_3 (x)}$ ($p_c = 4$), 0.91 with $\ket{\psi^3_3 (x)}$ ($p_c=2$) and 0.88 with $\ket{\psi^1_3 (x)}$ ($p_c = 0.2$) from the green-solid solution functions.  
}
\label{fig:2ODE}
\end{figure}
\end{widetext}
}

\subsection{ML-aided results}
Finally, we demonstrate ML-aided solutions for a few different DEs given by the QuVa algorithm in Fig.~\ref{fig:intro}. The coefficients of the target DEs are given by $\kappa_j$ ($j=0,1,2$) and a number of system qubits is three with six variational parameters. Initially, we randomly choose 600 sets of the variational parameters $\{ \bar{\lambda} \}$ in the CP and put them into the QP to extract expectation value data $\langle \hat{\cal O}^{tot} \rangle$ for each DE and each depth. Then, a rough landscape of the total expectation value is plotted with respect to six parameters individually. Afterwards, we perform the regression and optimisation schemes to select the next best variational parameters to refine the landscape of $\langle \hat{\cal O}^{tot} \rangle$. 

With updating the set of ML parameters in the middle of the iterations over the next 600 additional runs for each depth, the solution candidate functions are carefully chosen by $\langle \hat{\cal O}^{tot} \rangle \le p_c$, which is a small value for each depth. As we discussed in Appendix \ref{Append01}, the results of the total expectation values are used for confirming the necessary condition to be the solutions of the target PDEs. We here show that the condidate sets with a three-qubit system successfully follow the pattern of analytical solutions of different DEs over 1200 data for each depth. 

In Fig.~\ref{fig:2ODE}, we plot analytic solutions from 2O DE in 1D and ML-aid solutions from our QuVa PDE solver. In the left side of Fig.~\ref{fig:2ODE}, it shows analytical solution functions with $\{ \kappa_2, \kappa_1,\kappa_0, V_{max}\} =  \{1, -1, 8, 0 \}$ (top),  $\{1, 3,25, 32\}$ (middle) and  $\{1, 1, 40, 40\}$ with $\kappa_n = 500$ (bottom) and the boundary condition is $f(0)=f(1)$. Since the analytical solutions are in general not normalised, we simply find the function points $f(x_j)$ with $x_j=j/8$ ($j=0,1,2,...7$) and a normalisation condition is applied by $\sum_{j=0}^{7} |f(x_j)|^2 = 1$ corresponding to a discretised and normalised function. 

At the top of Fig.~\ref{fig:2ODE}, it turns out that the analytical solution functions of the no-potential 2O DE with different values of $f(0)>0$ (solid lines on the top left of Fig.~\ref{fig:2ODE}) become a single discretised function with the normalisation condition as shown in a blue curve at the top right side of Fig.~\ref{fig:2ODE} and the similar behaviours occur for $f(0)<0$ in the green curve of the top right of Fig.~\ref{fig:2ODE}. Interestingly, we see the tendency that reflected curves along the $x$-axis in the solutions between blue and green ones. For this simple 2O DE, the red dashed line shows one of ML-aid solver results in $\ket{\psi^2_3 (x)}$ with $\langle \hat{\cal O}^{tot} \rangle \le p_c=4$ from Eq.~(\ref{eq:Helm02}) and it approximately gives fidelity 0.92 with the discretised function with $f(0)<0$ (green solid line on the top right of Fig.~\ref{fig:2ODE}). For system ansatzs $\ket{\psi^d_3 (x)}$, a pattern of the solution function with $f(0)>0$ was not found in $\ket{\psi^d_3 (x)}$ ($d \le 3$) with high fidelity over 1200 data. It implies that it is difficult to predetermine the coefficient ${\cal C}_{000}$ corresponding to $f(0)$ in the variational ansatz.

In the middle of Fig.~\ref{fig:2ODE}, the figures are depicted with harmonic potential term $V(x)$ and $\{ \kappa_2, \kappa_1,\kappa_0, V_{max} \} = \{1, 3,25, 32 \}$. For the potential case, the left side of the figures are shown with different $f(0)$ and a significant dip is shown in the middle of the analytical curves due to the effect of the harmonic potential centred at $x=1/2$ with $V_{max} = 32$ in the region of $0\le x \le 1$. For its discretised version on the right side, it shows that the fidelity is roughly 0.91 between $\ket{\psi^3_3}$ in the red dashed line and $f(x)$ with $f(0) = -0.19$ in the green solid one ($p_c = 2$).

For the nonlinear case, the similar methods are applied with $\{ \kappa_2, \kappa_1,\kappa_0, V_{max}, \kappa_n \} = \{1, 1, 40, 40, 500\}$. The analytical curves show a rapid oscillation on the bottom left of Fig.~\ref{fig:2ODE} and its large nonlinearity brings a sensitive result depending on the value of $f(0)$. For fidelity, the bottom right curves show the fidelity approximately 0.88 between the ML-aid solution curve $\ket{\psi^1_3}$ and the discretised function with $f(0) = 0.54$ ($p_c = 0.2$). In general, the similar function patterns appear in different depths with large nonlinearity but it is highly limited to chase the rapid oscillating solutions with nine-point functions.

\section{Summary and further discussions}

In summary, we have described a ML-aided quantum variational (QuVa) solver for PDEs. It aims to provide a selected set of solution candidates to solve generalised PDEs. The main idea is that the expectation values of quantum operators give enough information about the solution functions of the target PDEs. The results are extracted from the data of measuring a few controlled qubits in the QPs with the support of ML techniques in the CP. We also demonstrated three examples of this solver for 1D 2O DE with a three-qubit ansatz and the fidelity is shown as higher than 0.88. One of the key advantages in this algorithm, we can recycle the expectation value data (e.g., $ \langle \hat{A}^{\dag} \rangle $ and $ \langle \hat{\rho} \rangle$) to find solutions of PDEs with different $\kappa_j$ because the landscapes of expectation values are independent from the PDE parameters.

For some nonlinear PDEs, the separation of variables may not be applicable \cite{NonPDE}. For instance, an entangled ansatz state in 2D will be considered as $\ket{\Psi(x,y)} \ne \ket{\psi(x)} \otimes  \ket{\phi(y)}$ in the QuVa method. 
In this case, we can keep utilising a generalised entangled ansatz $\ket{\Psi(x,y)}$ for Eqs.~(\ref{eq:2DLaplace01}) and (\ref{eq:2DLaplace02}) and additional conditional SWAP gates could be utilised for obtaining the expectation value $\bra{\Psi} \hat{\cal O}_{\partial^2} (x,y) \ket{\Psi}$ with highly entangled ansatzs in the QPs instead of using Eq.~(\ref{eq:2DExpect02}). In addition, there are many interesting nonlinear PDE problems in relativistic quantum dynamics and cosmology (e.g., Schr\"odinger-Newton equations \cite{Diosi84, Penrose2014,Howl}) that can be examined in the QuVa PDE solver.

Although the QuVa solver is designed to solve a general PDE form, we would like to mention a few obstacles to solve some target PDEs in the current QuVa approach.
From Eq.~(\ref{eq:Genreal_total_O}), this solver approach shows the limitation of finding a good solution to be $\langle \hat{O}^{tot} \rangle \approx 0$ because the range of expectation values are always $-1 \le \langle\hat{ A}^{\dag}\rangle , \,  \langle \hat{\cal V} \rangle, \, \langle \hat{\rho}_D \rangle \le 1$. For example, in the simple 2O DE with $\langle \hat{\cal V} \rangle = \langle \hat{\rho}_D \rangle = 0$, the total expectation value $\langle \hat{\cal O}^{tot} \, \rangle$ is limited between $\kappa_0$ and $\kappa_0 - {2 \over \delta L} \left( {2  \over \delta L} - { \kappa_1} \right) $. Thus, this approach may not provide an appropriate solution if $\kappa_0 < 0$ and $\kappa_1 \le {2\over \delta L}$ although the ML schemes will provide candidates wtih a minimum of $\langle \hat{\cal O}^{tot} \, \rangle$.

There are several interesting open questions beyond the scope of this paper based on the venue of computer sciences. We artificially fixed some crucial factors in the three-qubit demonstration (e.g.,how to choose the value of $p_c$ and what could be the optimal number of variational parameters and of the size of the initial data set in the QuVa solver). Thus, these investigations can guide us to prove the efficiency of the algorithm in the future. In addition, it will be important to study how to utilise the solution information of pre-learned data with the $N$-qubit system to investigate the solution of the other PDEs with a larger qubit system. 

\section*{Acknowledgements}
This work is supported by Basic Science Research Program through the National Research Foundation of Korea (NRF) funded by the Ministry of Education, Science and Technology (NRF-2021M3H3A1038085).
JJ would like to acknowledge J. Huh and D. K. Park for useful comments and discussions.

\appendix
\section*{Supplementary information}
\section{Basics of quantum information}
\label{BasicQI}
\subsection{Single- and two-qubit gates}
Mathematically, a qubit is a $2 \times 1$ column vector described by $\ket{0} = \left(
\begin{array}{c}
1 \\
0 \\
\end{array}
\right)$ and $\ket{1} = \left(
\begin{array}{c}
0 \\
1 \\
\end{array}
\right)$ \cite{NC_QIQC}. The two orthonormal vectors represent a quantum version of bit information similar to conventional information theory. To implement the arbitrary single qubit, we use three rotational operators such as $\hat{R}^{{\sigma}} (\lambda) = e^{-i \hat{\sigma}/2} = \cos {(\lambda / 2)} {\hat{I}} - i \sin {(\lambda / 2)} \, \hat{\sigma}$ for Pauli operators $\hat{\sigma}=\hat{X}, \, \hat{Y}, \, \hat{Z}$ such that $\hat{X} = \left(
\begin{array}{cc}
0 & 1 \\
1 & 0 \\
\end{array}
\right)$, $\hat{Y}= \left(
\begin{array}{cc}
0 & -i \\
i & 0 \\
\end{array}
\right)$ and $\hat{Z}= \left(
\begin{array}{cc}
1 & 0 \\
0 & -1 \\
\end{array}
\right)$.
For example, the rotational operators along one of the axes are given by
\begin{eqnarray}
\hat{R}^X (\lambda) &&=  \cos {(\lambda / 2)} {\hat{I}} - i \sin {(\lambda / 2)} \, \hat{X} ,  \label{R-X} \\
\hat{R}^Y (\lambda) &&=  \cos {(\lambda / 2)} {\hat{I}} - i \sin {(\lambda / 2)} \,\hat{Y} ,  \label{R-Y} \\
\hat{R}^Z (\lambda) &&=  \cos {(\lambda / 2)} {\hat{I}} - i \sin {(\lambda / 2)} \,\hat{Z} , \label{R-Z}
\end{eqnarray}
In addition, one of the important one-qubit gates is called a Hadamard gate $\hat{H}={1 \over \sqrt{2}} (\hat{X}+\hat{Z}) = i \hat{R}^Y (-{\pi \over 2}) \hat{R}^X (\pi)$.

For two-qubit gates, a CNOT gate $CX_{kl}$ is typically used for a two-qubit entangling gate on control qubit $k$ and target qubit $l$ such as
\begin{eqnarray}
CX_{kl}= \ket{0}_k\bra{0} \otimes \hat{I}_l + \ket{1}_k\bra{1} \otimes \hat{X}_l\, .
\end{eqnarray}
Note that there are many other alternatives of the CNOT gate for entangling two qubits in theory \cite{NC_QIQC}.

\subsection{$N$-qubit system states (ansatzs)}
\label{N-qubit ansatz states}

If one build a unit quantum gate $\hat{U}^{unit}$ consisting of single- and two-qubit gates with a set of variational parameters $\{ \bar{\lambda} \}$, the $d$-depth ansatz state with $N$ qubits is given by 
\begin{eqnarray}
\ket{\psi^d_N (\bar{\lambda})} = \left[\hat{U}^{unit} (\bar{\lambda}) \right]^d \hat{U}^{p} (\bar{\lambda})\, \ket{0}^{\otimes N}\,,
\label{eq:ansatz01}
\end{eqnarray}
where the parameter and unit operators are given by $\hat{U}^{p} (\bar{\lambda}) = \left[ \hat{R}^Y (\bar{\lambda}) \right]^{\otimes N}$ and 
\begin{eqnarray}
\hat{U}^{unit} (\bar{\lambda}) =  \Big( \hat{U}^{p} (\bar{\lambda}) \,  CX_{1,N} \prod_{j=1,..N-1} CX_{j+1,j} \Big).
 \label{eq:Append04} 
\end{eqnarray}
For example, the three-qubit ansatz states $\ket{\psi^d_3 (\bar{\lambda})}$ with three parameters $\bar{\lambda} =\{ \lambda_1, \lambda_2, \lambda_3 \}$ and depth-$d$ are given by
\begin{eqnarray}
&& \ket{\psi^{0}_3 (\bar{\lambda})}_{123}  = \Big( \hat{U}^{p} ({\lambda_1 , \lambda_2,\lambda_3})  \Big) \ket{000}_{123} ,~~~~~
\label{Depth0} \\
&& \ket{\psi^{d}_3 (\bar{\lambda})}_{123} =  \left[ \hat{U}^{unit} (\bar{\lambda})\right]^d  \ket{\psi^{0}_3 (\bar{\lambda})}_{123},
\label{Depth2} 
\end{eqnarray}
for $ \hat{U}^{unit} (\bar{\lambda})  = \Big( \hat{U}^{p} ( {\lambda_1 , \lambda_2 , \lambda_3})  \Big)  CX_{13} CX_{32} CX_{21} $ and  $\hat{U}^{p} ( {\lambda_1 , \lambda_2 , \lambda_3}) = \hat{R}^Y (\lambda_1) \hat{R}^Y (\lambda_2) \hat{R}^Y (\lambda_3) $.

In the main text, we use an ansatz state with six variational parameters $\bar{\lambda} =\{ \lambda_1, \lambda_2, \lambda_3 \lambda_4, \lambda_5, \lambda_6 \}$ and
\begin{eqnarray}
\hat{U}^{unit} (\bar{\lambda}) &=& \big( CX_{13} CX_{32} CX_{21}\big)  \hat{U}^{p} ({ \lambda_1, \lambda_2, \lambda_3})   \nonumber \\ && ~~~~~
 \big( CX_{13} CX_{32} CX_{21}  \big) \hat{U}^{p} ({ \lambda_4, \lambda_5, \lambda_6}).~~~~~~
 \label{eq:Append05} 
\end{eqnarray}

\section{Expectation value with a residual of PDEs} 
\label{Append01}
If $ \ket{\psi (\bar{x})}$ is an ansatz state of a nonlinear PDE, a quantum residual with non-invertible PDE matrix $\hat{{\cal O}}^{tot} =  \hat{{\cal O}}_\partial + \hat{{\cal O}}_{NL}$ 
for $\hat{\cal O}^{tot}  \, \ket{\psi (\bar{x})} = \ket{\delta}$ is given by
\begin{eqnarray}
&& Res_Q  (\bar{x}) = \bra{\psi (\bar{x})} \left( \hat{{\cal O}}^{tot}\right)^{\dag} \hat{{\cal O}}^{tot} \, \ket{\psi (\bar{x})} = \langle \delta \ket{\delta} , \label{Append_PDE01} \\ 
&& = \bra{\psi (\bar{x})}  
\Big( \hat{{\cal O}}_\partial^{\dag} \hat{{\cal O}}_{NL} 
+ \hat{{\cal O}}_{NL}^{\dag} \hat{{\cal O}}_\partial 
+ \hat{{\cal O}}_\partial^{\dag} \hat{{\cal O}}_{\partial} 
+ \hat{{\cal O}}_{NL}^{\dag} \hat{{\cal O}}_{NL} \Big) \, \ket{\psi (\bar{x})}.
\nonumber
\end{eqnarray}
Thus, the variational solution wavefunction $\ket{\psi (\bar{x})}$ should be satisfied $\ket{\delta} \approx \ket{\vec{0}}$ and $Res_Q  (\bar{x}) \approx 0$ in Eq.~(\ref{Append_PDE01}).

\begin{widetext}

\begin{figure}[t]
\centering
\includegraphics[width=0.36\linewidth,trim=1.5cm 0cm 0cm 0cm]{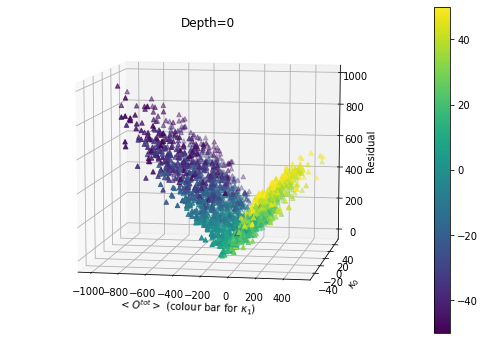}
\includegraphics[width=0.36\linewidth,trim=0cm 0cm 1.5cm 0cm]{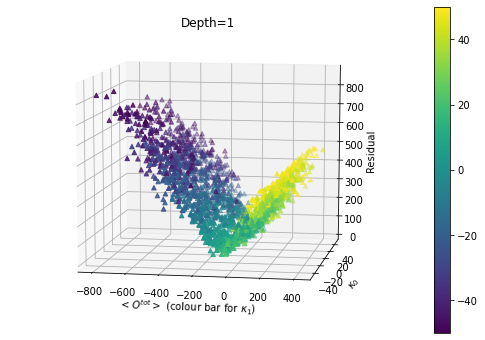}
\includegraphics[width=0.36\linewidth,trim=1.5cm 0cm 0cm 0cm]{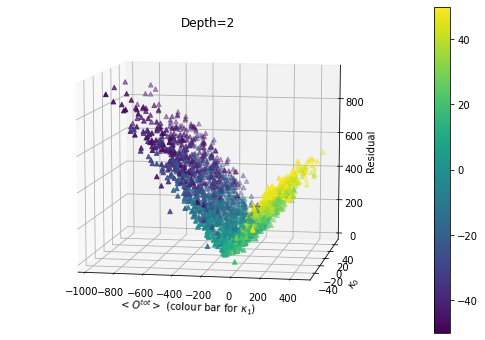}
\includegraphics[width=0.36\linewidth,trim=0cm 0cm 1.5cm 0cm]{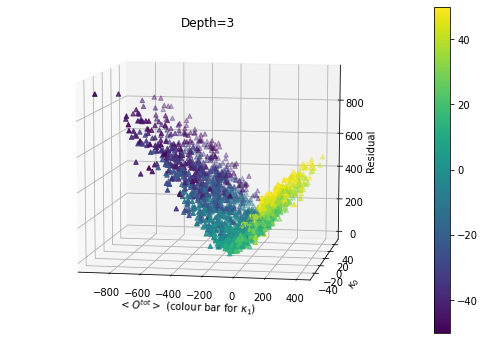}
\caption{The evidence of the necessary condition between $\langle \hat{\cal O}^{tot}  (\bar{\lambda}) \rangle$ and $Res (\bar{\lambda})$ with $\kappa_2 = 1$ and $-50 \le \kappa_j \le 50$ ($j=0,1$).}
 \label{fig:correlation_residual}
\end{figure}

\end{widetext}

In an alternative viewpoint, we add $ \ket{\psi (\bar{x})}$ in the both side of Eq.~(\ref{General_Quan_PDE01}) such as 
\begin{eqnarray}
\left( \hat{{\cal O}}^{tot} + \hat{I}\right) \,\ket{\psi (\bar{x})} =\ket{\psi (\bar{x})} + \ket{\delta}
\label{Append_PDE01-1}
\end{eqnarray}
for , and
\begin{eqnarray}
&& \bra{\psi (\bar{x})} \left( \hat{{\cal O}}^{tot} + \hat{I} \right)^{\dag} \left( \hat{{\cal O}}^{tot} + \hat{I}\right) \, \ket{\psi (\bar{x})} \nonumber  \\
&& ~~~~~~~~~~~~~~=  1 +  \langle \psi (\bar{x}) \ket{\delta} + \langle \delta \ket{\psi (\bar{x})} + \langle \delta \ket{\delta}.~~~~~~~~~~
\label{Append_PDE02}
\end{eqnarray}
Thus, the left side of the equation is given by
\begin{eqnarray}
Res_Q  (\bar{x})  + \bra{\psi ({\bar{x}})} 
\left( \hat{{\cal O}}_\partial^{\dag} + \hat{{\cal O}}_{NL}^{\dag} + \hat{{\cal O}}_\partial + \hat{{\cal O}}_{NL} + \hat{I} \right) \, \ket{\psi ({\bar{x}}) }. \nonumber \\
&& \label{Append_PDE03}
\end{eqnarray}
From Eq.~(\ref{Append_PDE02}), this becomes
\begin{eqnarray}
\Re \left[ \bra{\psi ({\bar{x}})}\hat{{\cal O}}^{tot} \ket{\psi ({\bar{x}}) } \right] = \Re \left[ \langle \delta \ket{\psi (\bar{x})}  \right]\,.
\label{Append_PDE04}
\end{eqnarray}
Therefore,  $\langle \hat{\cal O}^{tot}  (\bar{x}) \rangle \approx 0$ if either $ \ket{\delta} \approx \ket{\vec{0}}$ or $ \ket{\delta}$ becomes perpendicular to $\ket{\psi (\bar{x})}$.
We tested this necessary condition between $\langle \hat{\cal O}^{tot}  (\bar{\lambda}) \rangle$ and $Res_Q (\bar{\lambda})$ with $\ket{\psi^d_3 (x)} $ for 2O DE. Figure \ref{fig:correlation_residual} shows the strong correlation evidence between them on the same $\ket{\psi^d_3 (x)} $ in Eq.~(\ref{eq:Expect05}) regardless of its depth ($d=0,1,2,3$).

\section{Expectation value calculator for a quantum system}
\label{General-expectation-calculation}
We here explain how to calculate the expectation value of a specific operator $\hat{\cal O}$, which is made of quantum circuits, and how to efficiently obtain it using the statistics of a single controlled qubit (called qubit $C$) \cite{Ekert_Oi}. The detail protocol is given as below.

\begin{enumerate}
\item Initialisation: Initialise a qubit in $\ket{0}_C$ (named control qubit) and $N$ qubits in $\ket{0}^{\otimes N}$ (also named system qubits). The control qubit state is prepared in $\ket{+\phi}_C = e^{-i\phi/2} \hat{R}^Z (\phi) \, H\, \ket{0}_C$.

\item System preparation: Prepare the system state $\ket{\psi^d_N (\bar{\lambda}) }$ with depth $d$ on system qubits. The total gate in Eq.~(\ref{eq:ansatz01}) is performed on the system qubits with the parameter set $\bar{\lambda} = \{ \lambda_j | j=1, ..., k \}$ ($k$ is the number of variational parameters).

\item Operator preparation: Perform a controlled-unitary gate $CU_{CS}$ based on a designed quantum operator $\hat{\cal O}_S$ between $\ket{+}_C$ and $\ket{\psi^d_N}_S$ such as $\ket{\Psi^{d}_N}_{CS} $ = $CU_{CS} \ket{+\phi}_C \ket{\psi^d_N}_S$. Note that  $CU_{CS} = \ket{0}_C \bra{0} \otimes \openone_S +  \ket{1}_C \bra{1} \otimes \hat{\cal O}_S$.

\item Measurement: Measure only the control qubit in $Z$-axis after a Hadamard gate on the control qubit.

\item Data collection: Repeat the protocol from 1 to 4 to obtain faithful statistical data of either $\langle  \hat{Z} \rangle$ = $\Re \left[ \bra{\psi^d_N} \hat{{\cal O}}_S \ket{\psi^d_N} \right]$ with $\phi =0$ or $\langle \hat{Z} \rangle$ = $\Im \left[ \bra{\psi^d_N} \hat{{\cal O}}_S \ket{\psi^d_N} \right]$ with $\phi=-{ \pi\over 2}$.
\end{enumerate}

\section{Preparation of desired multi-qubit states}
\label{Overlap_fidelity}
We here would like to show how to construct a desired multi-qubit state as a mixed state. In particular, what we aim to create is a specific mixed state (diagonalised density matrix) to be utilised for calculating both potential and nonlinear terms in our QuVa PDE solver.  

\begin{figure}[t]
\centering
 \includegraphics[width=0.95\linewidth,trim=2cm 2.5cm 2cm 2cm]{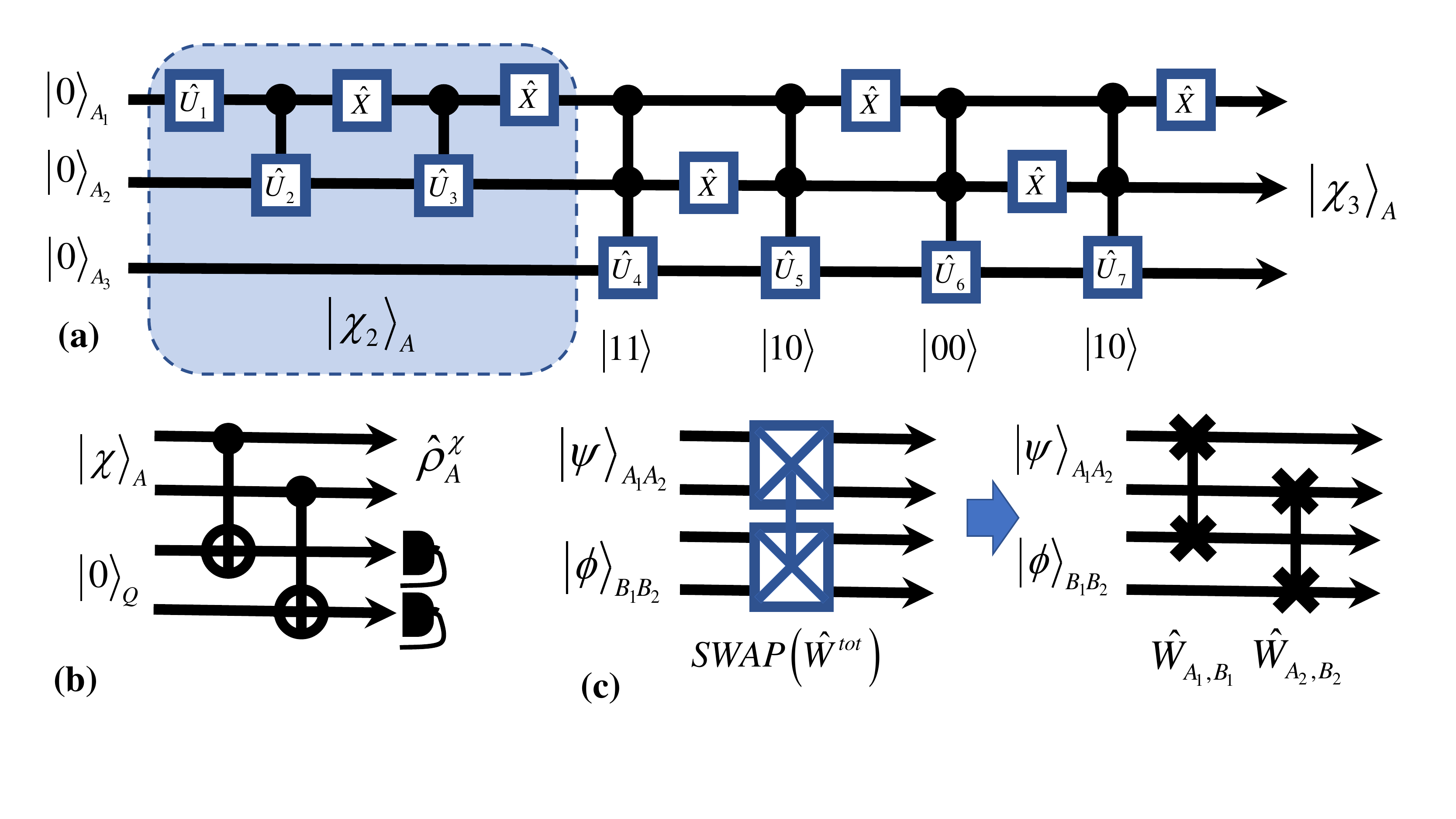}
\caption{(a) A quantum circuit for building a desired $N$-qubit state $\ket{\chi_N}$. In the blue box, $\ket{\chi_2}$ is made by two controlled unitary gates with two $\hat{X}$ gates and a unitary gate $\hat{U}_1$. Then, four controlled-controlled unitary gates are mainly used to build target three-qubit state $\ket{\chi_3}$ from $\ket{\chi_2}$ and show that individual single-qubit gates ($\hat{U}_4$ to $\hat{U}_7$) are applied to the corresponding states from $\ket{11}_{A_1 A_2}$ to $\ket{10}_{A_1 A_2}$. (b) It shows how to build desired diagonal matrix $\hat{\rho}^{\chi}$ from $\ket{\chi}$ using CNOT gates and measurements. (c) A block-SWAP gate $\hat{W}^{tot}$ between two quantum systems consists of sequential SWAP gates such as $\hat{W}^{tot} = \hat{W}_{A_1,B_1}\hat{W}_{A_2,B_2}$.
}
\label{fig:01}
\end{figure}

As shown in Fig.~\ref{fig:01}(a), it is well-known that an arbitrary single-qubit can be made of (at least) two rotational gates such that $ \ket{\chi_1}_{A_1} = \hat{U}_1 \ket{0}_{A_1} = a_0 \ket{0} + a_1 \ket{1}$. For two-qubit states, the second qubit in $\ket{0}_{A_2}$ is added on $\ket{\chi_1}_{A_1}$. Then, two controlled unitary gates ($\hat{U}_2$ and $\hat{U}_3$) with two NOT gates ($\hat{X}$) make the desired two-qubit state $\ket{\chi_2}$ given by 
\begin{eqnarray}
\ket{\chi_2}_{A} && = \left( \hat{X}_1  CU_3  \hat{X}_1 CU_2 \right) \ket{\chi_1}_{A_1}  \ket{0}_{A_2}  \nonumber \\
&& = a_{00} \ket{00}_{A} +  a_{01} \ket{01}_{A} +  a_{10} \ket{10}_{A} +  a_{11} \ket{11}_{A}.~~~~~~~
\label{Arbit-qubit01}
\end{eqnarray}
Thus, $\ket{\chi_2}_{A_1 A_2} = a_{00} \ket{00} + a_{01} \ket{01} + a_{10} \ket{10} + a_{11} \ket{11}$ as desired two-qubit states. As shown in Fig.~\ref{fig:01}(a), a desired three-qubit state $\ket{\chi_3}_{A}$ is built by four additional controlled-controlled unitary gates with four NOT gates.

\subsection{Artificial decoherence process}
\label{Arti-decoh}
This is a quantum processor for constructing desired mixed state $\hat{\rho}^{\chi}_A$ from pure state $\ket{\chi}_A$.
One of the universal methods to make a density matrix form (as a diagonal matrix) is to perform artificial decoherence process given by ancillary qubits in $Q$. For two-qubit states, let us assume that the pure state $\ket{\chi_2}$ in Eq.~(\ref{Arbit-qubit01}).

 As shown in Fig.~\ref{fig:01} (b), we simply apply pairwise CNOT gates between a qubit in $A$ and a qubit in $Q$ and the total four-qubit state before the measurement is given by 
\begin{eqnarray}
\ket{D^{tot}} &&= CNOT_{A_1, Q_1}  CNOT_{A_2, Q_2} \ket{\chi_2}_{A_1 A_2}\ket{00}_{Q_1 Q_2}  \nonumber \\
&& =a_{00} \ket{0000} +  a_{01} \ket{0101}+  a_{10} \ket{1010} +  a_{11} \ket{1111}. \nonumber \\
&& 
\end{eqnarray}
Once we measure all the extra qubits in $Q$ and ignore the results of the qubits, the final outcome state is given by
\begin{eqnarray}
\hat{\rho}^{\chi} && = \sum_{j,k=0,1} {}_{Q_1 Q_2} \bra{jk}{D^{tot} } \rangle \bra{D^{tot} } {jk}\rangle_{Q_1 Q_2}  \nonumber \\
&& = \sum_{j,k=0,1} |a_{jk}|^2 \ket{jk}_{A_1 A_2} \bra{jk}, 
\end{eqnarray}
for $j, k = 0,1$.
Therefore, this mixed state is now prepared as a desired diagonal matrix in $A$ and used for calculating both potential and nonlinear terms in Fig.~\ref{fig:03} (a).

\subsection{Overlap fidelity calculator for two quantum systems}
In addition to construct a mixed state in the QP, a SWAP operation between two quantum systems plays a key role in the QuVa PDE solver. A single-qubit SWAP gate is given by
\begin{eqnarray}
\hat{W}_{A,B} = \sum_{j,k=0}^{1} \ket{j}_A\bra{k} \otimes \ket{k}_B \bra{j},
\end{eqnarray}
\cite{NC_QIQC} and two qubits are swapped through the operator $\hat{W}$. For example, if $ \ket{\chi}_A = a_0 \ket{0}_A + a_1 \ket{1}_A$ $\ket{\eta}_B = b_0 \ket{0}_B + b_1 \ket{1}_B$, the SWAP operated state is given by
\begin{eqnarray}
\hat{W}_{A,B} \ket{\chi}_A \ket{\eta}_B = \ket{\eta}_A  \ket{\chi}_B. 
\label{SWAP}
\end{eqnarray}

As shown in Fig.~\ref{fig:01}(c), we can swap two quantum systems as a block-SWAP gate $\hat{W}^{tot}$.
If we combine the block-SWAP gate between two-qubit states ($\ket{\chi}_{A}$ and $\ket{\eta}_{B}$) and with the expectation value calculator, the statistical results of the single-qubit measurement in $C$ is given by
\begin{eqnarray}
&& \langle \hat{W}_{A,B} \rangle = \bra{\chi}_{A} \bra{\eta}_{B} \left( \hat{W}_{A,B} \right) \ket{\chi}_{A} \ket{\eta}_{B} = \left| \langle \chi \ket{\eta}\right|^2,~~~~
\label{Expect-SWAP}
\end{eqnarray}
where $\ket{\chi}_{A} = \sum_j a_j \ket{j}$ and $\ket{\eta}_{B} = \sum_k b_k \ket{k}$.
Therefore, based on the expectation value opeation with pure state $\ket{\chi}_{A}$ and mixed state $\hat{\rho}= \sum_g \rho_g \ket{g}\bra{g}$ in $B$, the statistical results of the control-qubit measurement in $C$ is equal to 
\begin{eqnarray}
\langle \hat{W}_{A,B} \rangle = \bra{\chi}\, \hat{\rho} \, \ket{\chi}.
\label{Expect-SWAP-mixed}
\end{eqnarray}

\end{document}